*Yu-Jun Cui,[1] Anh-Minh Tang,[2] Duilio Marcial,[3] Jean-Marc Terpereau, [4] Gilles Marchadier, [4] Xavier Boulay, [2]*

# Use of a differential pressure transducer for the monitoring of soil volume change in cyclic triaxial test on unsaturated soils

---

**ABSTRACT:** A new experimental set-up using a differential pressure transducer was developed, that enables the monitoring of volume changes in cyclic triaxial tests on unsaturated soils. Calibration tests were performed in order to analyze the performance of the set-up, especially in terms of loading frequencies. Based on calibration results, a low frequency of 0.05 Hz was adopted for the tests carried out on the unsaturated loess from northern France. Five water contents were considered in the tests. The obtained results have confirmed the efficiency of the new system for volume change monitoring under cyclic loading. The effect of water content on the cyclic behavior of loess was clearly evidenced. Finally, some suggestions were made to improve the accuracy of the system.

## Introduction

During the rainy season 2001-2002, several sinkholes were observed along the TGV (French high-speed train) line connecting Paris and Lille, France. In order to understand these stability problems, a geotechnical characterization of the loess deposits of northern France was carried out (Cui et al. 2005), and the cyclic behavior of this unsaturated loess was studied by performing cyclic triaxial tests. This kind of tests needed developing an appropriate volume change monitoring technique. Indeed, in triaxial drained tests on saturated soil, the volume change of the sample is usually monitored by pore-water volume exchange. However, for unsaturated soils, the total volume change of the specimen is no longer equal to the pore-water volume change; therefore the conventional method is no longer applicable. Local displacement transducers are usually used to measure the local vertical and radial strains of soil specimens (Blatz and Graham 2000; Kawai et al. 2002; Cunningham et al. 2003; Sun et al. 2004; Ibraim and Di Benedetto 2005; Cabarkapa and Cuccovillo 2006). The limitation of this method is that it is generally applicable for rigid specimens; that is obviously not the case of most natural soils. The electro-optical laser system developed by Romero et al. (1997) has the advantage of no direct contact between specimen and measurement device, but it needs sophisticated installation procedures. For these reasons, the volume change of unsaturated soils in triaxial tests is usually monitored by following cell fluid exchanges.

[1] Professor, Ecole Nationale des Ponts et Chaussees, Paris, (CERMES) – Institut Navier, 6 et 8 av. Blaise Pascal, Cite Descartes, Champs-sur-Marne, 77455 MARNE-LA-VALLEE CEDEX 2, France. Email: cui@cermes.enpc.fr.
[2] Ecole Nationale des Ponts et Chaussees.
[3] Universidad Central de Venezuela, Caracas, Venezuela.

The volume change of the confining fluid in a triaxial cell under loading is the sum of: (i) the soil specimen volume change; (ii) the volume changes of the fluid, the cell, and the tubing etc. Agus et al. (2000), Sultan et al. (2002), and Tang et al. (2006) used stainless steel cell and tubing in order to minimize the system's volume change. The use of such a system needs, however, careful calibration in terms of the effects of pressure, temperature, especially in the low pressure range. The difficulty related to the volume change of the system can be overcome by using double walled cell technique. Bishop and Donald (1961) added an inner cell filling with mercury in the conventional triaxial cell that was filled with water. As the two cells were connected, changes in the confining pressure do not result in any deformation of inner cell. The soil volume change can be then deduced from the change of the fluid level in the inner cell. This level, monitored by a cathetometer, can be indicated by a ball floater (Bishop and Donald 1961), a target float (Matyas and Radhakrishna 1968), or a thin layer of silicon oil if mercury is replaced by water (Cui and Delage 1996). Water volume change in the inner cell can also be monitored using other techniques as paraffin-water burette (Wheeler 1988), electric volumeter (Yin 2003) or volume-pressure controller (Estabragh et al. 2004). Rampino et al. (1999), Aversa and Nicotera (2002), and Ng et al. (2002) used a differential pressure transducer (DPT) to monitor the level change of the fluid inside the inner cell.

[4] French Railway Company (SNCF), France.

The different techniques presented above were applied for monotonic shear test. As far as cyclic shear testing is concerned, obviously, the method using cathetometer can not be applied because it operates manually and it is difficult to follow the cyclic movement of fluid level; On the contrary, the technique using DPT seems promising. That is the reason why this technique was adopted in the present study. It should be noted that DPT is usually used for the monitoring of volume change in monotonic shear test, and in the knowledge of the authors, it has never been used in a cyclic triaxial test. In this paper, a triaxial apparatus which enables cyclic loading tests on unsaturated soils with volume monitoring using DPT is presented. Five tests were carried out using this apparatus on the loess from northern France at various water contents. The obtained results clearly evidenced the effect of water content on the cyclic behavior of the loess, and moreover, they showed the relevance of the experimental set-up for cyclic testing on unsaturated soils.

## Testing devices

### *Triaxial cell*

A DPT was installed in a cyclic triaxial cell for monitoring the total volume change of unsaturated soil specimen. The schematic layout of the system is presented in Figure 1. The vertical pressure applied on soil specimen is controlled by the difference between pressures applied in the two chambers of the vertical loading system. The loading controller, that also controls the confining pressure, monitors these pressures. A pressure transducer is installed below the base pedestal to monitor the vertical stress applied on the specimen. The vertical displacement is monitored by a LVDT (Linear Variable Differential Transformer). All the data including applied pressures, vertical stress and vertical displacement are recorded by the logging system through the control system. Details on the loading control and the vertical displacement measurement of this cyclic triaxial cell were described by De Gennaro et al. (2004).

This cyclic cell was modified and adapted to volume change monitoring when testing unsaturated soils. The principle of double walled triaxial cell was applied by adding an inner cell in plexiglas. As water inside and outside the inner cell is under the same confining pressure, any variation of the water level inside the inner cell, indicated by the DPT changes corresponds to a volume change of the soil specimen. A preliminary calibration is required for data treatment.

In cyclic triaxial tests, the measurement of soil volume change requires a very high accuracy. For this reason, a low working pressure, high resolution DPT was used. Moreover, the response time of DPT must be compatible with the frequency at which the cyclic loading is applied during a test. The specification of the DPT is presented in Table 1. The full-scale pressure output of the DPT is 1250 Pa that corresponds to a water height of 125 mm. The accuracy is ±1.25 Pa (0.125 mm water height). The response time under a pressure of 1125 Pa is less than 100 milliseconds. Therefore, theoretically, the DPT is able to work under a cyclic pressure of a frequency lower than 10 Hz.

Table 1. Specifications of the differential pressure transducer.

| Specification | Value |
| --- | --- |
| Full scale pressure output (Pa) | 1250 |
| Accuracy (Pa) | 1.25 |
| Resolution (Pa) | 0.025 |
| Response time | < 100 milliseconds for 1125 Pa |
| Frequency response | Flat to 10 Hz |

*Calibration*

The DPT was first calibrated with static pressure, and the following relationship was found: $p = 245.75\Delta U$, where $p$ is the pressure in Pa and $\Delta U$ is the output variation of DPT in V. It should be noted that this calibration was established in the positive range of pressure; however, in a real test, the pressure could be in the negative range. It can be the case when fixing the initial differential water level between outside and inside the inner cell; it can be also the case during the unloading phase. Tarantino and Mongiovi (2003) calibrated a commercial pressure transducer in the negative range, and they observed the same calibration coefficient for the positive and negative ranges. In the present work, it is assumed that the relationship above can be applied for the negative range of pressure.

The response of the DPT under cyclic loading with various frequencies was first studied. The cell was assembled without soil specimen. Five frequencies were considered: 0.05 Hz, 0.1 Hz, 0.2 Hz, 0.5 Hz and 1 Hz; two displacements of piston were accounted for: ± 3 mm and ±15 mm. The results are shown in Figure 2. In each case with a given frequency and a given displacement of piston, the records from LVDT for piston displacement (mm) and from DPT for volume change (V) are presented. The calculated vertical displacement of piston ($\Delta h_{cal}$) from DPT records is compared with the recorded one ($\Delta h_p$), and the comparison is also shown in the figure. $\Delta h_{cal}$ was calculated using Eq. (1):

$$\Delta h_{cal} = \frac{S_c - S_p}{S_p} \Delta h_w = \frac{S_c - S_p}{S_p} \times \frac{249.75}{g \times \rho_w} \times \Delta U \tag{1}$$

where

$S_c$ is the cross section of inner cell (100 mm in diameter), $S_c$ = 78.54 cm$^2$;

$S_p$ is the cross section of piston (48.6 mm in diameter), $S_p$ = 18.55 cm$^2$;

$\Delta h_w$ is the variation of water level in the inner cell, in m;

$g$ is the gravity acceleration, $g$ = 9.81 m/s$^2$;

$\rho_w$ is the volumetric mass of water, $\rho_w$ = 1002 kg/m$^3$ at 20 °C (after Incropera and De Witt 1996);

$\Delta h_{cal}$ is the calculated displacement of piston, in m.

In test (a), the piston moved within ±15 mm at a frequency of 0.05 Hz by following a sinusoidal function of time. It can be observed that both the LVDT and DPT gave a sinusoidal output signal. As the response time of the DPT was less than 0.1 second, no delay is visible between the vertical displacement of piston ($\Delta h_p$) and the DPT output ($\Delta U$) by comparing $\Delta h_p - t$ and $\Delta U - t$ plots. A slight delay was however evidenced in $\Delta h_p - \Delta h_{cal}$ plot: the relationship $\Delta h_p - \Delta h_{cal}$ is of a thin ellipse shape. This delay means when $\Delta h_p$ reached its maximum value, $\Delta U$ had not yet.

The same phenomenon can be observed in test (b) where the piston moved within ±3 mm with a frequency of 0.05 Hz. As the frequency in tests (a) and (b) was the same, the shape of the ellipse in the $\Delta h_p - \Delta h_{cal}$ plan is similar. In test (c) where the piston displacement was kept at ±3 mm and the frequency was increased to 0.1 Hz, it can be observed that the ellipse becomes larger. That means the delay between $\Delta h_p$ and $\Delta h_{cal}$ increased. In test (d), a higher frequency of 0.2 Hz was applied. It was observed that the LVDT gave satisfactory response since maximum displacement value (±3 mm) was always reached. Concerning the $\Delta h_p - \Delta h_{cal}$ relationship, a larger delay was observed between both.

In test (e) and test (f), higher frequencies were applied: 0.5 Hz and 1 Hz for tests (e) and (f) respectively. It can be observed that the monitoring system fails in such high frequencies. Indeed, in test (e), the LVDT was not able to correctly follow the piston displacement within ±3 mm; the out put signal of LVDT and DPT is no longer sinusoidal; the $\Delta h_p - \Delta h_{cal}$ plot shows an extremely important delay between both. The situation was worse in test (f): the piston displacement was not correctly followed by the LVDT, neither in shape nor in amplitude; the DPT gave meaningless output; there was no rational $\Delta h_p - \Delta h_{cal}$ relationship. It can be then concluded that the capacity of the monitoring system is limited at a loading frequency lower than 0.5 Hz.

A linear regression relation $\Delta h_p = \alpha \Delta h_{cal}$ was attempted to be made for tests (a) to (e). The results of α and the determination coefficient ($R^2$) are presented as a function of frequency in Figure 3. It can be observed that the correlation between $\Delta h_p$ and $\Delta h_{cal}$ is the best in the case of test (b) with ±3 mm displacement and 0.05 Hz frequency. When the amplitude is more important, ±15 mm in test (a), the DPT gave a slightly underestimated response ($\alpha < 1$). In case of frequencies higher than 0.05 Hz, even at small amplitude of ±3 mm, it appeared that the increase of frequency induced a quasi-linear decrease of α and $R^2$. That means, higher is the frequency, more is the DPT underestimation and worse is the correlation.

In conclusion, the calibration tests performed and presented above show that the DPT could be used to monitor the water level change in the inner cell when the rate of this change is low enough. In the case of test (b), ±3 mm displacement at 0.05 Hz frequency, the calculated displacement ($\Delta h_{cal}$) was well correlated with the measured one ($\Delta h_p$) ($\alpha = 1.0099$ and $R^2 = 0.98$). On the other hand, this good correlation also validated the calculation using geometric dimensions of the system. That enables the estimation of the total volume change of soil specimen during a cyclic shear test. The total volume decrease of soil specimen ($\Delta V$) generated by a downward displacement ($\Delta h_p$) of the piston can be then calculated using the water level rise ($\Delta h_w$) inside the inner cell measured by the DPT (Eq. (2)):

$$\Delta V = \Delta h_p S_p - \Delta h_w \left(S_c - S_p\right) \tag{2}$$

**Material and experimental program**

The soil tested is a loess taken from Northern France in the region of Picardie at a depth of 2.2 m, in blocs of a dimension of 40×40×60 cm. The blocs were conserved in plastic box and waxed on their surfaces in order to maintain its natural water content. The geotechnical properties of this soil are presented in Table 2. More details on the geological and geotechnical properties of this soil are described by Cui et al. (2004). It is observed that the soil is characterized by low plasticity (PI = 6), low dry density ($\rho_d = 1.39 Mg/m^3$), low degree of saturation ($S_r$ = 53%), low clay fraction (% < 2µm = 16), high carbonate content (12%) and low suction (34 kPa, determined using filter paper method, see Delage and Cui 2000). The low plasticity and low suction is correlated with the low clay fraction; the high carbonate content is typical of loessic soil. The presence of carbonates makes the soil a naturally cemented material with, therefore, a relatively high strength. This explains the low density of natural loessic soils.

Table 2. Properties of loess.

| Properties | Value |
|---|---|
| Water content (%) | 18.6 |
| Particle density ($Mg/m^3$) | 2.714 |
| Liquid limit (%) | 28 |
| Plastic limit (%) | 22 |
| Plasticity index | 6 |
| Dry density ($Mg/m^3$) | 1.39 |
| Saturation degree (%) | 53 |
| Carbonate content (%) | 12 |
| Suction (kPa) | 34 |
| Clay (< 2 μm) fraction (%) | 16 |

Prior to cyclic shear tests, soil specimens (70 mm in diameter, 140 mm high) were prepared by trimming. Air dry method and wetting method using wetted filter paper were applied to control the water contents of specimens.

Cyclic shear tests were carried out at five different water contents: 10, 14, 21, 23, and 27%. After mounting the triaxial cell, a confined pressure of 25 kPa was applied on the soil specimen. The drainage outlets were let open. Firstly, a cyclic deviator stress $q = 10 \pm 10$ (kPa) was applied at a frequency of 0.05 Hz for 700 cycles; and then, the maximum deviator stress is raised 10 kPa every 100 cycles until the failure of soil.

## Experimental results

Figure 4 presents the results obtained from the first five cycles of the test at 14% water content. After the application of the confining pressure (25 kPa), the deviator stress (q) was varied in a sinusoidal function of time between 0 and 20 kPa with a frequency of 0.05 Hz. A cyclic axial strain ($\varepsilon_1$) can be observed during this cyclic loading. Its amplitude (±0.03%) corresponds to a vertical displacement of ± 0.042 mm measured by the vertical displacement transducer (LVDT). It can be noticed that the sinusoidal function is not as net as expected for axial strains. This is because the maximum axial displacement recorded is close to the resolution of the LVDT (0.01 mm) on one hand and on the other hand, the results would be affected by any noise in data logging and control devices. From the $q-\varepsilon_1$ plot, it can be seen that the relationship between $q$ and $\varepsilon_1$ can be correlated with a linear function, $\mathrm{d}q/\mathrm{d}\varepsilon_1 = 40$ MPa.

The signal recorded by the DPT for volumetric strain ($\varepsilon_v$) is worse: it is not possible to distinguish any cycles in the $\varepsilon_V - N$ plot. This is related to the accuracy that the measurement system can provide. With a total soil volume of 538.8 $cm^3$, and a surface area between the inner cell and the piston $(S_c - S_p)$ of 60 $cm^2$, the water level accuracy measured by the DPT (±0.125 mm) corresponds to a volumetric strain accuracy of ±0.14%, which is larger than the maximum $\varepsilon_V$ value obtained, about 0.12%. Moreover, the volumetric strain is calculated from the data recorded by the DPT and the LVDT. The accuracy on the volumetric strain therefore depends on both the accuracy of DPT and that of LVDT. To which one must add the effects of any noise in data logging system and control devices.

The results of five cycles at the end of the test are shown in Figure 5. During these cycles, the deviator stress (q) was varied from 0 to 120 kPa following a sinusoidal function. The results obtained in terms of axial strains ($\varepsilon_1$) show a smooth cyclic function. This is due to the fact that an amplitude of ±0.15% which corresponds to ±0.21 mm in vertical displacement is distinctly higher than the accuracy of the LVDT (±0.01 mm). The volumetric strain measured varies with an amplitude of 0.15%. This value is similar to the accuracy of the DPT (±0.14% in $\varepsilon_v$). The secant shear modulus can be calculated from the $q - \varepsilon_1$ plan, $\mathrm{G} = \mathrm{d}q / \mathrm{d}\varepsilon_1 = 40$ MPa. This parameter did not change after 1700 cycles.

Figure 6 presents the results of the test at 14% water content from the beginning until the failure. During the first 700 cycles, few variations were observed on both volumetric strain ($\varepsilon_v$) and axial strain ($\varepsilon_1$). Significant axial strain was produced when the deviator stress (q) increased progressively from 0-20 kPa to 0-130 kPa. From the $q - \varepsilon_1$ plot it can be observed that the cyclic loading generated accumulated plastic axial strain. The failure observed at the cycle N = 17010 is characterized by an abrupt increase of both volumetric strain and axial strain. Just before the failure, the axial strain ($\varepsilon_1$) was 10% and the volumetric strain ($\varepsilon_v$) was 7%. The maximum volumetric strain obtained in this test was approximately 25%, which corresponds to a decrease of volume of 134600 $mm^3$. Considering the surface area $(S_c - S_p)$ of 60 $cm^2$ between the inner cell and the piston, this volume change should generate a water level change of 22.4 mm. This water level change (22.4 mm) is much lower than the capacity of the DPT (125 mm).

The results of five tests at various water contents are gathered in Figure 7. In this figure, the deviator stress (q), axial strain ($\varepsilon_1$), and volumetric strain ($\varepsilon_v$) are presented as a function of the cycle number (N). In $q - N$ plan, it can be seen that the deviator stress varied from 0 to 20 kPa during the first 700 cycles. And then, its peak value increased 10 kPa every 100 cycles until failure for all the tests except that at 10% water content where after 2000 cycles (q = 0-150 kPa), the peak value of deviator stress was increased 10 kPa every 10 cycles until failure at q = 240 kPa.

The effect of the water content was clearly observed in both $\varepsilon_1 - N$ and $\varepsilon_v - N$ plan. Before failure, at a same cycle number, the axial strain and volumetric strain were larger at higher water content. Moreover, except the case of 21% water content where the failure was not reached, the number of cycle, N, required to reach the failure decreased with the increase of water content. In the examining of the effect of the amplitude of q on changes in $\varepsilon_1$ and $\varepsilon_V$, it can be observed that the strains started generally increasing significantly and then seemed to stabilize under a given cyclic deviator stress. In addition, higher was the water content, larger the first part of quick change. An exception was made by 10% water content at which no effect of deviator stress amplitude was observed: both $\varepsilon_1$ and $\varepsilon_V$ varied linearly as a function of N.

**Discussion**

The results of calibration tests showed that the DPT could monitor, with a satisfactory accuracy, the cyclic displacement of piston within ±3 mm (that corresponds to a variation of ±0.93 mm of water level in the inner cell) at a frequency of 0.05 Hz. When the frequency increased, the DPT tended to give responses of underestimation. The DPT failed in monitoring when the frequency reached 1 Hz. In fact, in case of 1 Hz loading frequency, the response time of DPT (0.1 s) is not short enough to be neglected in cyclic test. The failure for frequencies higher than 0.05 Hz is because the interaction between water and metallic wall (water adhesion) which do not allow the water movement to follow the imposed piston frequency. Thus, it is believed that in case of important volume change (more compressible tested soils), the inner cell diameter can be larger, the adhesion effect would be less significant and higher frequencies could be applied.

The tests performed on soil specimens enabled the verification of DPT capacity. It was observed that the DPT could not follow the cyclic volumetric strains smaller than 0.14%. Nevertheless, since the total volume change of soil specimens is generally much larger during a full shear test until failure, the DPT can be used to monitor the volume change satisfactory. It was the case for the tested loess. The maximum volumetric strain obtained from the test at 14% water content was $\varepsilon_v$ = 25% that corresponds to a water rise of 22.5 mm in the inner cell. This value is still far from the capacity of the DPT (125 mm). On the other hand, the accuracy of the DPT (±0.125 mm of water height) corresponds to a volumetric strain accuracy of ±0.14%. This accuracy is not high enough for the monitoring of small cyclic strains. One of the methods to improve the accuracy of the system is to reduce the surface area between the inner cell and the piston ($S_c - S_p$). Decreasing this surface area would induce a more significant change in water level. When performing monotonic shear tests using DPT as monitoring technique, Ng et al. (2002), Aversa and Nicotera (2002) tried to minimize the surface area of the top of the inner cell ($S_c$). In the present case, the accuracy can be improved by increasing the section of the piston ($S_p$) in the measurement zone. For example, with a piston of 78 mm diameter ($S_p$ = 49.69 cm$^2$), the accuracy of the DPT in measuring the soil volume change would be five time higher (±0.03%). In that case, the recorded maximum soil volumetric strain (25%, that corresponds to 112.5 mm of water) would still not exceed the DPT capacity. Obviously, one can also improve the accuracy by using a DPT with a smaller working pressure, therefore higher accuracy.

As far as the behavior of the tested loess is concerned, it has been observed that the shear strength under cyclic triaxial loading is decreasing with water content increase. This phenomenon is in agreement with the experimental results obtained by running suction controlled shear tests on unsaturated soils (Cui and Delage 1996, Sun et al. 2004, Cunningham et al. 2003, Rampino et al. 2000, among others). Indeed, for an unsaturated soil, its shear strength is decrease by suction decreases or water content increases. In terms of soil volume changes, Cui and Delage (1996) observed that the volumetric strain was larger at lower suction. This is consistent with the water effect identified on the studied loess. The effect of the amplitude of cyclic deviator stress on volume changes at different water contents can be explained by an easier microstructure arrangement therefore a larger volumetric strain, when water content was higher.

## Conclusions

A differential pressure transducer (DPT) was used for the measurement of unsaturated soil volume change in a cyclic triaxial cell. Calibration tests performed at various frequencies showed the performance of the system at 0.05 Hz. Five tests at various water contents were carried out at this frequency. The tests performed using this system on the loess taken from Northern France showed that the cyclic volumetric strain could be satisfactorily monitored when its amplitude is higher than the accuracy of the DPT, ±0.14%. Improvements of the accuracy can be made by decreasing the surface area between the inner cell and the piston or by taking a DPT of smaller working pressure.

As far as the behavior of the studied loess is concerned, it was observed that cyclic shear strength was decreased by water content increase. This is consistent with the well-known behavior of unsaturated soils evidenced by monotonic shear tests with suction control.

## Acknowledgments

The authors are grateful to Réseau Ferré de France (Railway Network of France) for its financial support.

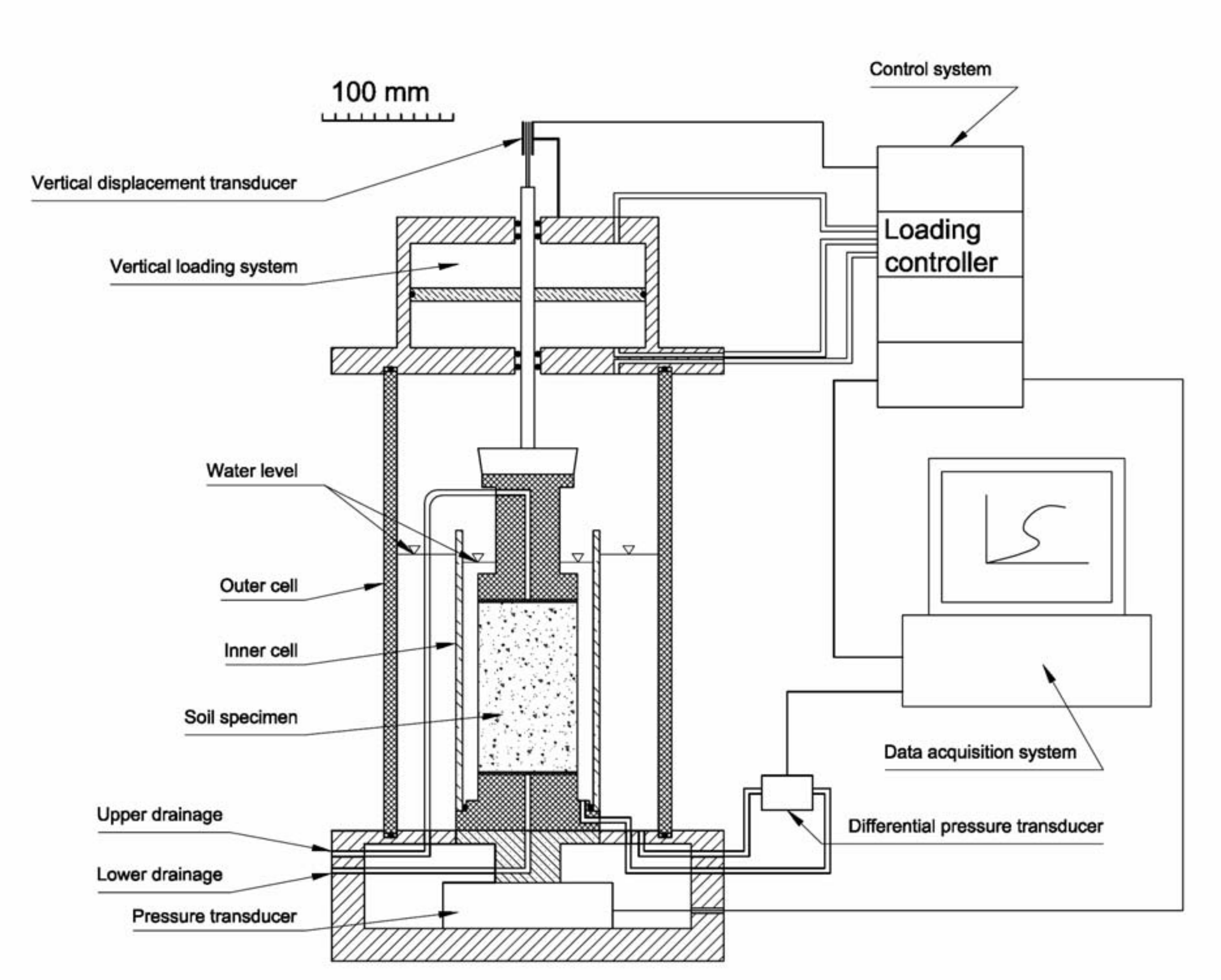

Figure 1. Cyclic triaxial cell equipped with a differential pressure transducer for total volume change monitoring.

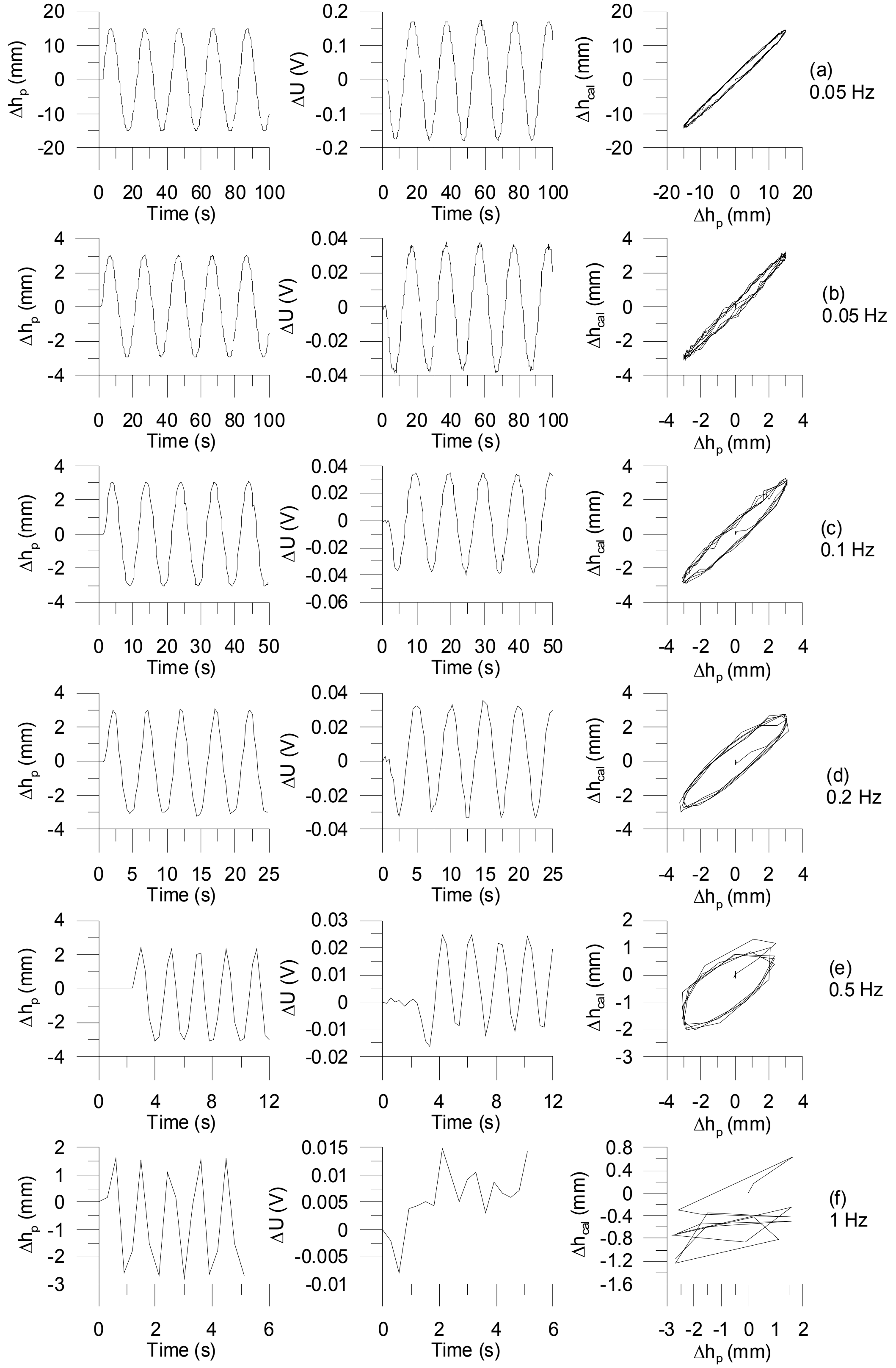

Figure 2. Results obtained from calibration tests with various frequencies.

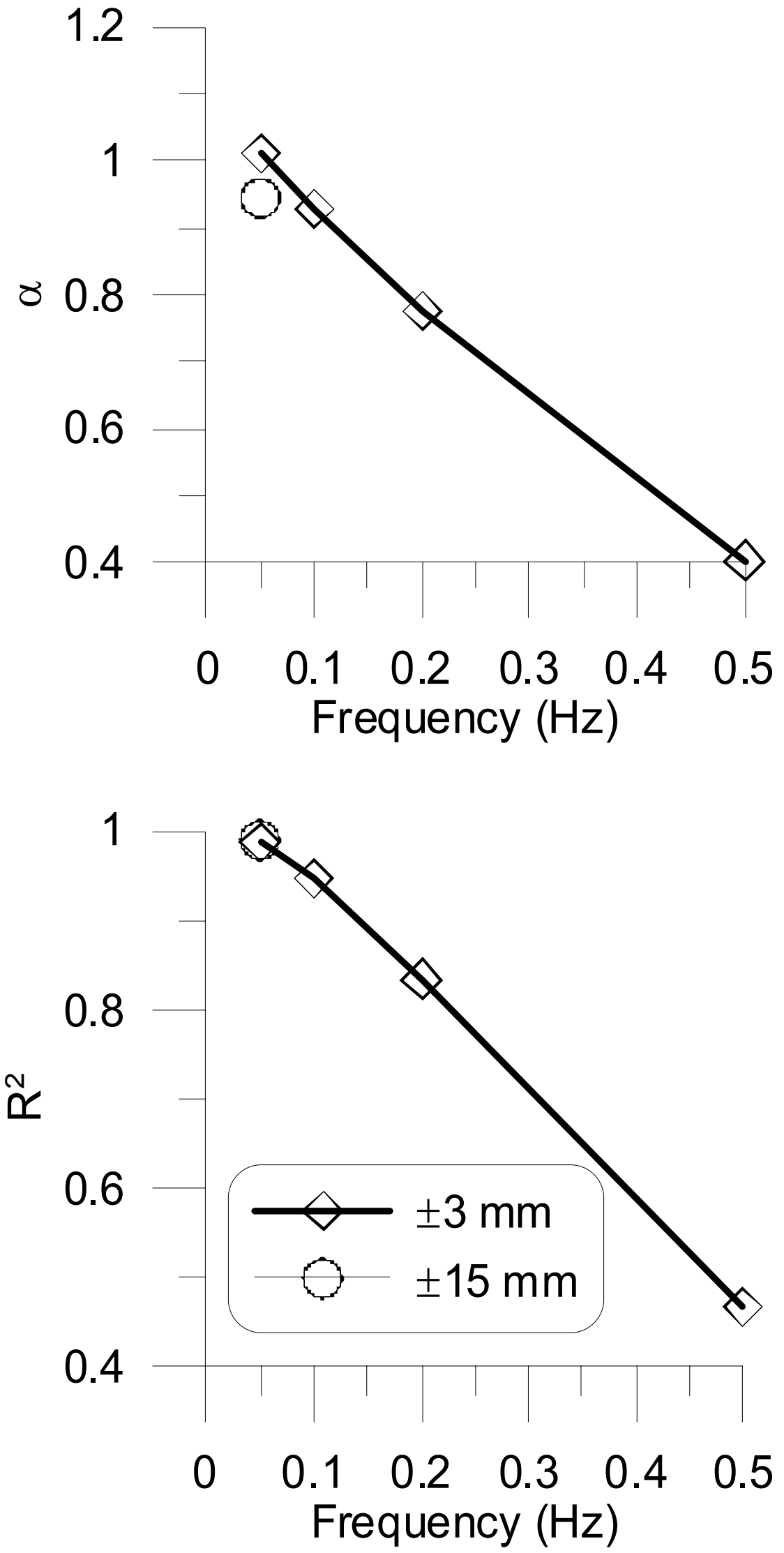

Figure 3. Evaluation of correlation between calculated displacement and measured displacement.

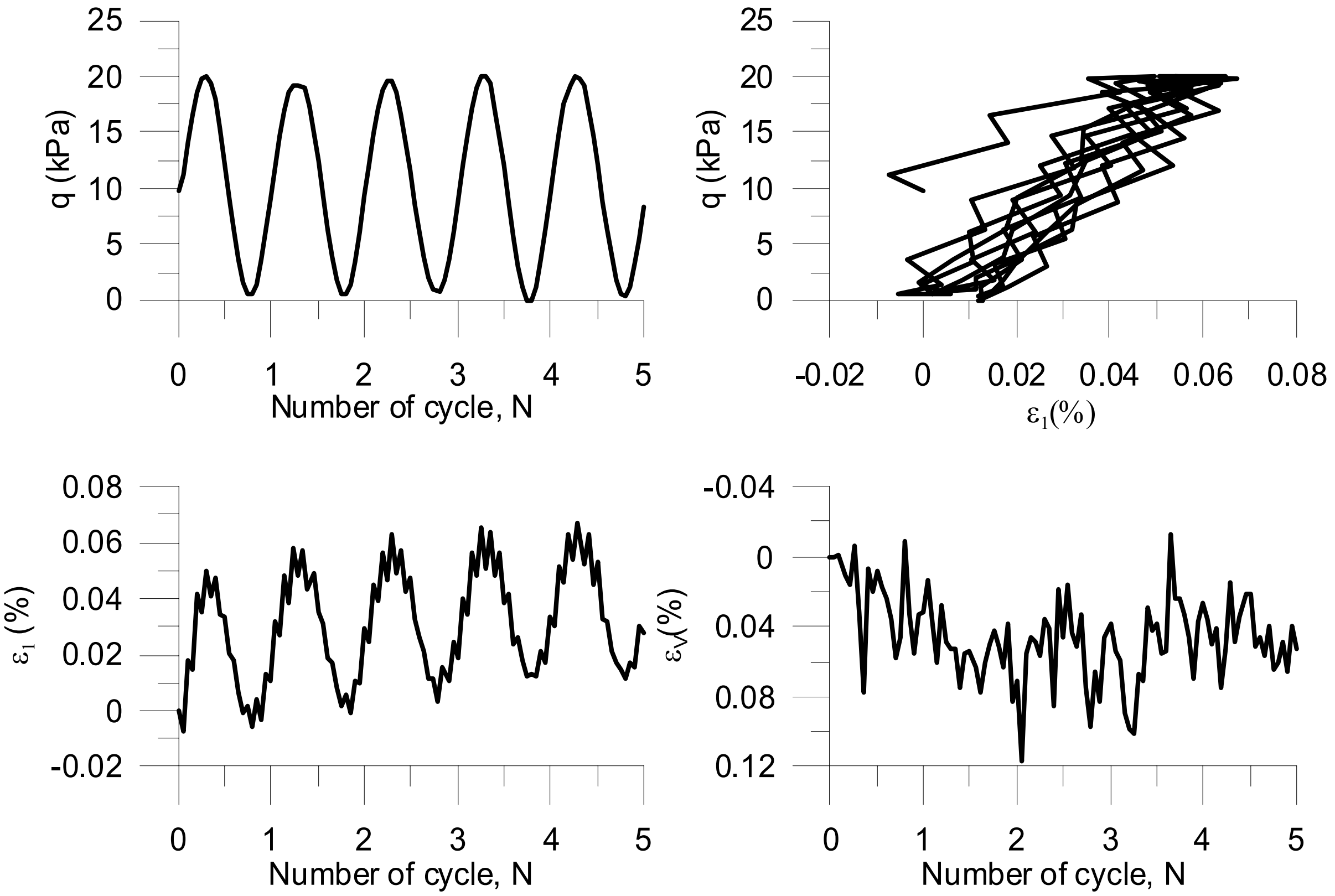

Figure 4. Cyclic triaxial test on unsaturated loess. w = 14 %, f = 0.05 Hz, $\sigma_3$ = 25 kPa. First cycles.

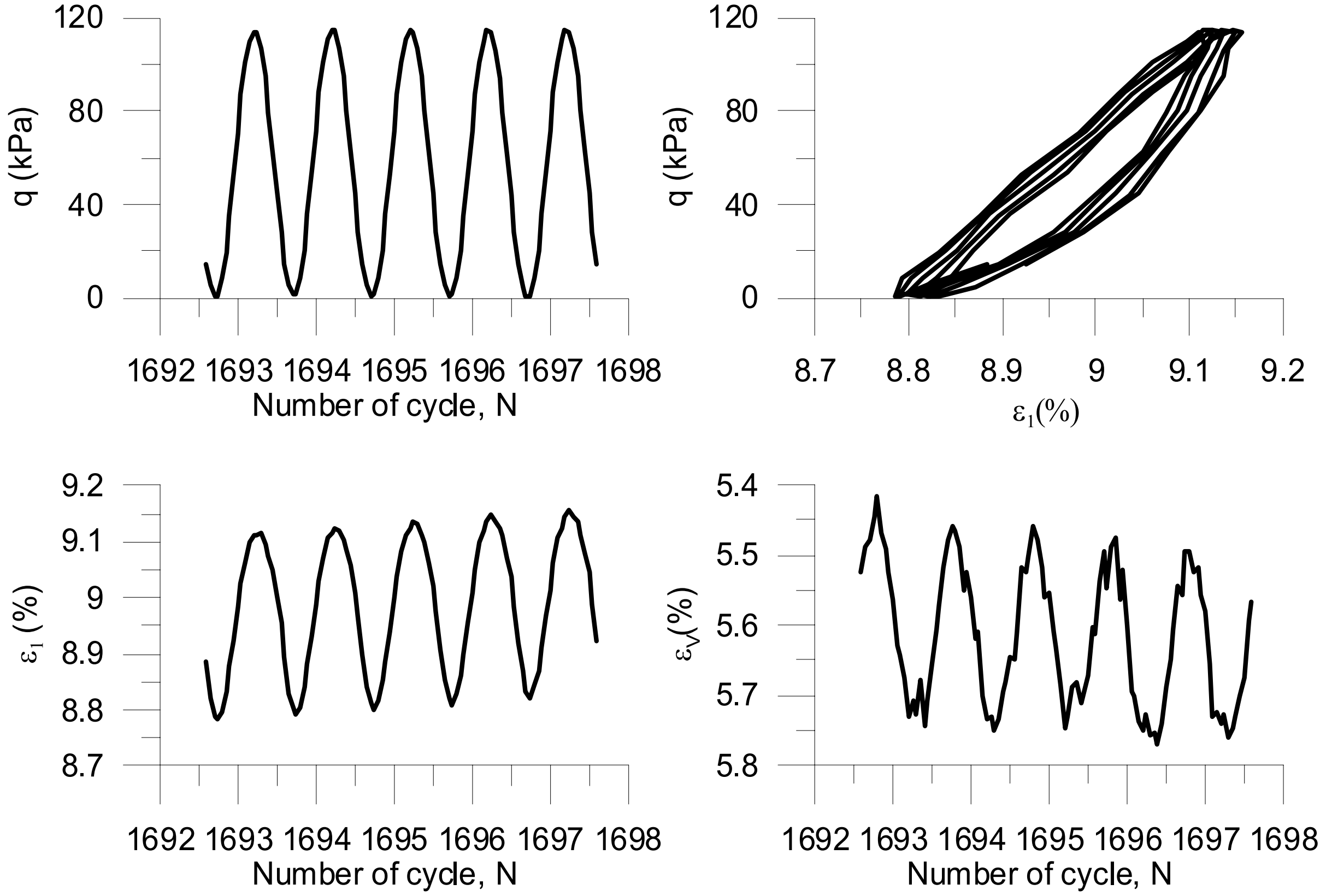

Figure 5. Cyclic triaxial test on unsaturated loess. w = 14 %, f = 0.05 Hz, $\sigma_3$ = 25 kPa. Last cycles.

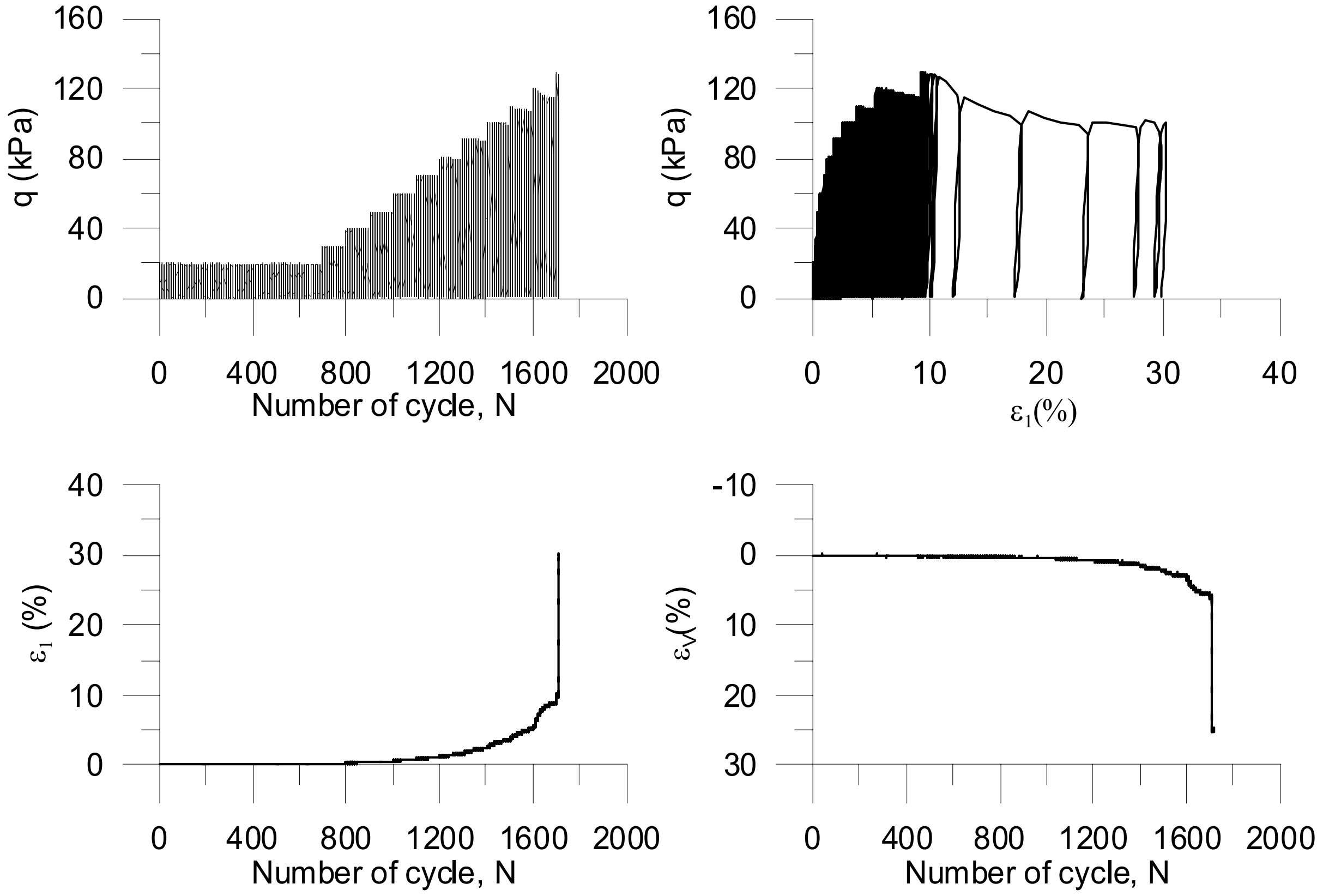

Figure 6. Cyclic triaxial test on unsaturated loess. w = 14 %, f = 0.05 Hz, $\sigma_3$ = 25 kPa.

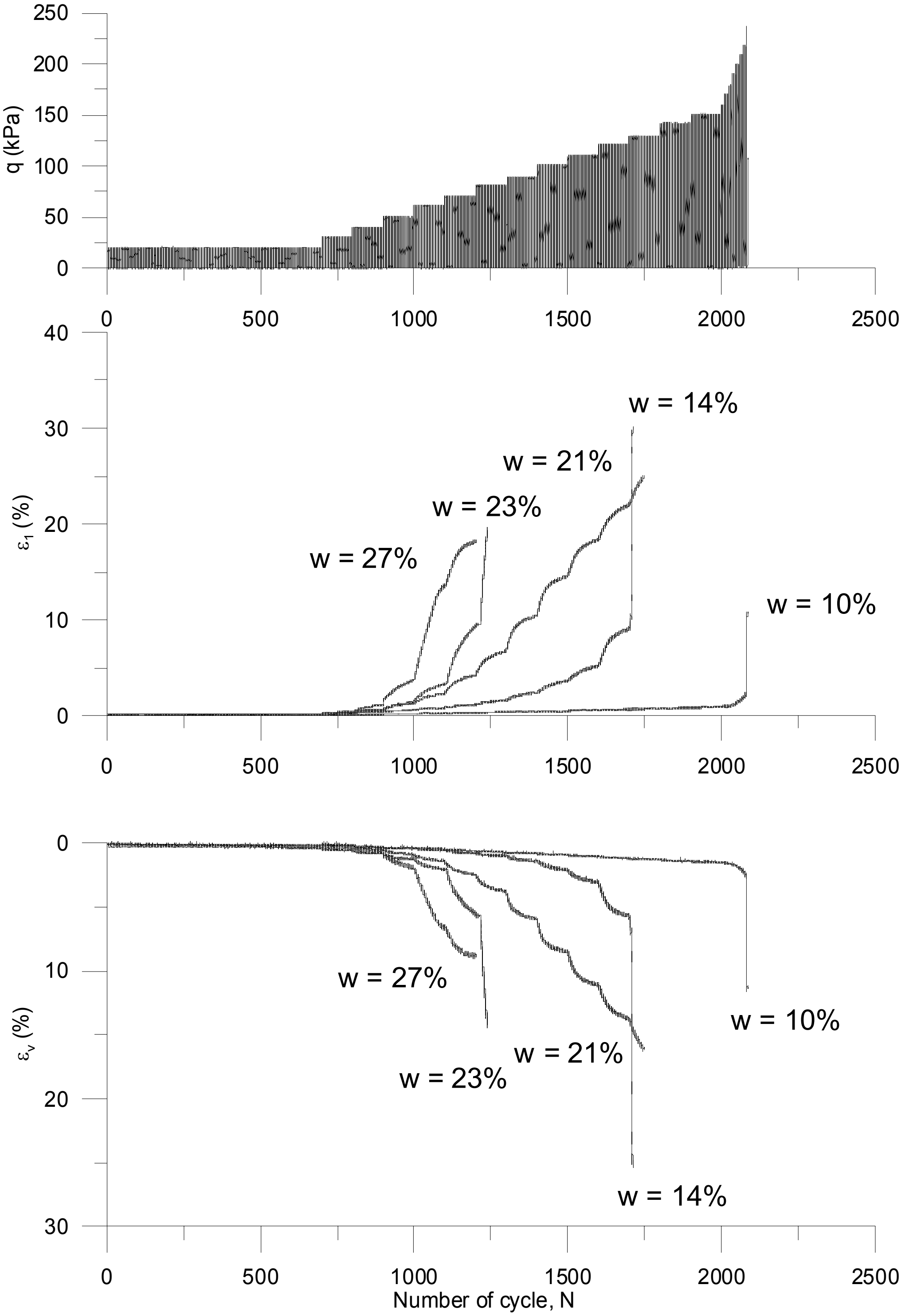

Figure 7. Cyclic triaxial tests on unsaturated loess at various water contents. f = 0.05 Hz, $\sigma_3$ = 25 kPa.